\documentclass[aps,pra,twocolumn,amsmath,amssymb,longbibliography,superscriptaddress]{revtex4-2}
\pdfpageattr {/Group << /S /Transparency /I true /CS /DeviceRGB>>}
\usepackage{graphicx}
\usepackage{color}
\usepackage[hidelinks]{hyperref}
\usepackage{orcidlink}
\usepackage{amsmath}
\usepackage{amsthm}
\usepackage{physics}
\usepackage{amsfonts}
\usepackage{times,txfonts}
\usepackage{lipsum}
\usepackage{bm}  
\usepackage{latexsym} 
\usepackage{amssymb}
\usepackage{comment}
\usepackage{array}      

\usepackage{txfonts}
\usepackage{xcolor}
\usepackage{float}
\usepackage{nameref}
\usepackage{placeins}
\hypersetup{colorlinks=true,linkcolor=blue,citecolor=blue,urlcolor=blue}

\usepackage[normalem]{ulem}

\begin{document}

\author{Dara Murphy\,\orcidlink{0009-0007-0151-7983}}
\email[]{dara.murphy@ucdconnect.ie}
\affiliation{School of Physics, University College Dublin, Belfield, Dublin 4, Ireland}
\affiliation{Rinn Quantum and Centre for Quantum Engineering, Science, and Technology, University College Dublin, Dublin 4, Ireland}

\author{George Mihailescu\,\orcidlink{0000-0002-0048-9622}}
\affiliation{School of Physics, University College Dublin, Belfield, Dublin 4, Ireland}
\affiliation{Rinn Quantum and Centre for Quantum Engineering, Science, and Technology, University College Dublin, Dublin 4, Ireland}

\author{Anne Matthies\,\orcidlink{0000-0003-1155-1281}}
\affiliation{School of Physics, University College Dublin, Belfield, Dublin 4, Ireland}
\affiliation{Rinn Quantum and Centre for Quantum Engineering, Science, and Technology, University College Dublin, Dublin 4, Ireland}
\affiliation{Department of Physics, Harvard University, Cambridge, Massachusetts 02138, US}

\author{Irene D'Amico\,\orcidlink{0000-0002-4794-1348}}
\address{School of Physics, Engineering and Technology, University of York, York, United Kingdom}
\address{York Centre for Quantum Technologies, University of York, York, United Kingdom}

\author{Steve Campbell\,\orcidlink{0000-0002-3427-9113}}
\affiliation{School of Physics, University College Dublin, Belfield, Dublin 4, Ireland}
\affiliation{Rinn Quantum and Centre for Quantum Engineering, Science, and Technology, University College Dublin, Dublin 4, Ireland}

\title{Energetic channels in finite many-body quantum charging}


\begin{abstract}
We examine how a finite many-body system can function as an effective charger for a single-qubit quantum battery, using the Lipkin–Meshkov–Glick model as a paradigmatic setting. As our figure of merit, we consider the battery ergotropy while explicitly resolving the energetic flows across the composite system. In the absence of intra-charger interactions, we show that the generation of non-zero ergotropy is controlled by the competition between collective coupling and spectral mismatch. We further demonstrate that even passive positive-temperature charger states can generate finite battery ergotropy, provided the charger is sufficiently large and sufficiently hot. Intra-charger interactions restructure the charger spectrum and qualitatively modify the available energetic pathways. While they can enhance the amount of energy accessible for charging, they can simultaneously divert energy into internal many-body and battery–charger interaction channels, producing an energetic impedance to direct charger-to-battery transfer. Consequently, battery ergotropy alone is insufficient to characterize an effective many-body charger: the useful work generated and its energetic provenance must be assessed together.

\end{abstract}
\date{\today}
\maketitle

\section{Introduction}

Quantum thermodynamics has grown substantially over the last two decades into a broad framework for understanding work, heat, irreversibility, and control at microscopic scales, where coherence, entanglement, finite-size effects, and environmental coupling can modify thermodynamic behaviour in ways absent from the macroscopic limit~\cite{Campbell2026,Goold2016}. Within this broader programme, it has become increasingly important to understand how energy is stored, transported, and released when genuinely quantum effects are at play, motivating the rapidly growing field of quantum batteries~\cite{Campaioli2024,Ferraro2026, Alicki2013, Francica2020,Pekola2021, Murphy2025,Simon2025,cheng2026,hoang2026,xuan2026,Rinaldi2026}. Quantum batteries provide a useful setting for gaining foundational insight into how thermodynamic notions must be reformulated for finite quantum systems~\cite{Campbell2026, Campaioli2024}, while also being increasingly proposed as practical microscopic work reservoirs, for example, in emerging quantum technologies~\cite{Kurman2026}.

Following Alicki and Fannes~\cite{Alicki2013}, a quantum battery can be viewed as a \(d\)-dimensional quantum system that stores energy and subsequently releases work through a controlled time-dependent perturbation of its Hamiltonian. In this setting, the relevant energetic figure of merit is not simply the total energy stored in the device, but rather the fraction that can be extracted as useful work. This is the ergotropy, defined as the maximum work that can be extracted from a quantum state by cyclic unitary operations~\cite{Allahverdyan2004},
\begin{align}
\mathcal{E}(\hat{\rho})
&=
\operatorname{Tr}\!\left[\hat{\rho}\hat{H}\right]
-
\min_{\hat{U}}
\operatorname{Tr}\!\left[
\hat{U}\hat{\rho}\hat{U}^{\dagger}\hat{H}
\right]
\nonumber\\
&=
\sum_{j,k} r_j \epsilon_k
\left(
\left|\langle r_j|\epsilon_k\rangle\right|^2
-\delta_{jk}
\right),
\label{eq:Ergotropy}
\end{align}
where the minimization is taken over all unitary transformations \(\hat U\), for which the global solution is known: \(r_j\) and \(\epsilon_k\) denote the eigenvalues of \(\hat{\rho}\) and \(\hat H\), respectively, ordered such that \(r_1\ge r_2\ge\cdots\) and \(\epsilon_1\le\epsilon_2\le\cdots\), while \(\lvert r_j\rangle\) and \(\lvert\epsilon_k\rangle\) are their corresponding eigenstates. Ergotropy has therefore naturally become a central concept in the theory of quantum batteries, providing an operational measure of the useful energy stored in a quantum system, with a growing body of studies dedicated to examining the extraction of work from a variety of systems~\cite{Hovhannisyan2020,Biswas2022,feliu2024,Hadipour2025,Razzoli2025,grazi2026,Beder2026}, including exploring non-universality of loads for quantum batteries~\cite{ostberg2026charging}. Recently, the subtle distinction between energy and ergotropy has been explored, for example, in the context of their transport efficiency~\cite{Tirone2021,Simon2025,Murphy2025,Beder2026}. 

Building on this framework, substantial effort has been devoted to identifying many-body platforms capable of enhancing quantum-battery performance. Collective coupling in Dicke-type batteries can strongly accelerate charging, and the role of many-body interactions within such architectures has also been investigated~\cite{Farrero2018,Campaioli2024,Andolina2019,Farre2020,yan2026}. Genuine quantum advantage has been identified in, for example, complex SYK and anharmonic bosonic batteries~\cite{Rossini2020, Andolina2025, fox2024harnessing}. A broad parallel literature has explored interacting spin systems, including cavity-driven, dissipative and noisy settings, and has established that internal interactions, collective structure and environmental couplings can substantially modify charging power, stored energy and work extraction~\cite{Zhao2025,Zhang2024,Evangelakos2025,Dou2022,Bhattacharya2026,Le2018,Zhao2021,yang2020,Huangfu2021,Arjmandi2022,romero2025}. Related studies have examined critical quantum batteries~\cite{Barra2022,Ho2026,Donelli2025,Yadav2026}, central-spin architectures~\cite{Liu2021,Peng2021,ShiPRB2025,ShiPRB2026}, and collectively interacting many-body systems~\cite{Li2025,Yadav2026,shukla2026collectivedynamicsversusentanglement,shukla2026manybody}.

The ability of a battery system to gain ergotropy via an auxiliary charger has been studied in a variety of contexts and models~\cite{Andolina2018, Farina2019, Huangfu2021, Peng2021, Liu2021, Joshi2022}, often restricting attention to interactions that conserve energy and excitations. This makes the energetic origin of the charging process relatively transparent. In this work, we relax this constraint. The central question motivating the present work is therefore: what determines whether a finite many-body system functions as an effective quantum charger?
Here we address this question using the Lipkin--Meshkov--Glick (LMG) model as a finite many-body charger for a central qubit battery. Originally introduced in nuclear physics~\cite{LIPKIN1965188,MESHKOV1965199,GLICK1965211}, the LMG model has since become a canonical fully connected spin system, with its collective and critical properties studied extensively across many-body physics~\cite{Ma2009,Castanos2006,champion2025}. This richness has also made it, and closely related collective many-body settings, useful for exploring a range of critical behavior as well as quantum-batteries, including investigations of enhanced charging and many-body constraints on charging performance~\cite{Farre2020,abah2022,Seidov2024,Ho2026,Caneva2008,wang2026}.

We show that whether a finite many-body system functions as an effective charger depends non-trivially on the available excitation sectors, system size, spectral alignment with the battery, internal interactions and temperature. We establish that a sufficiently hot positive-temperature charger, despite being passive with respect to its own Hamiltonian, can still generate local battery ergotropy. By carefully assessing the associated energy flows, we further demonstrate that achieving large battery ergotropy does not, in and of itself, imply that the many-body system has functioned effectively as a charger; rather, detuning and intra-charger interactions can redistribute energy through the battery–charger coupling, obscuring the energetic origin of the charging process.


\section{Preliminaries}

\begin{figure}[tb]
    \centering
    \includegraphics[width=0.9\linewidth]{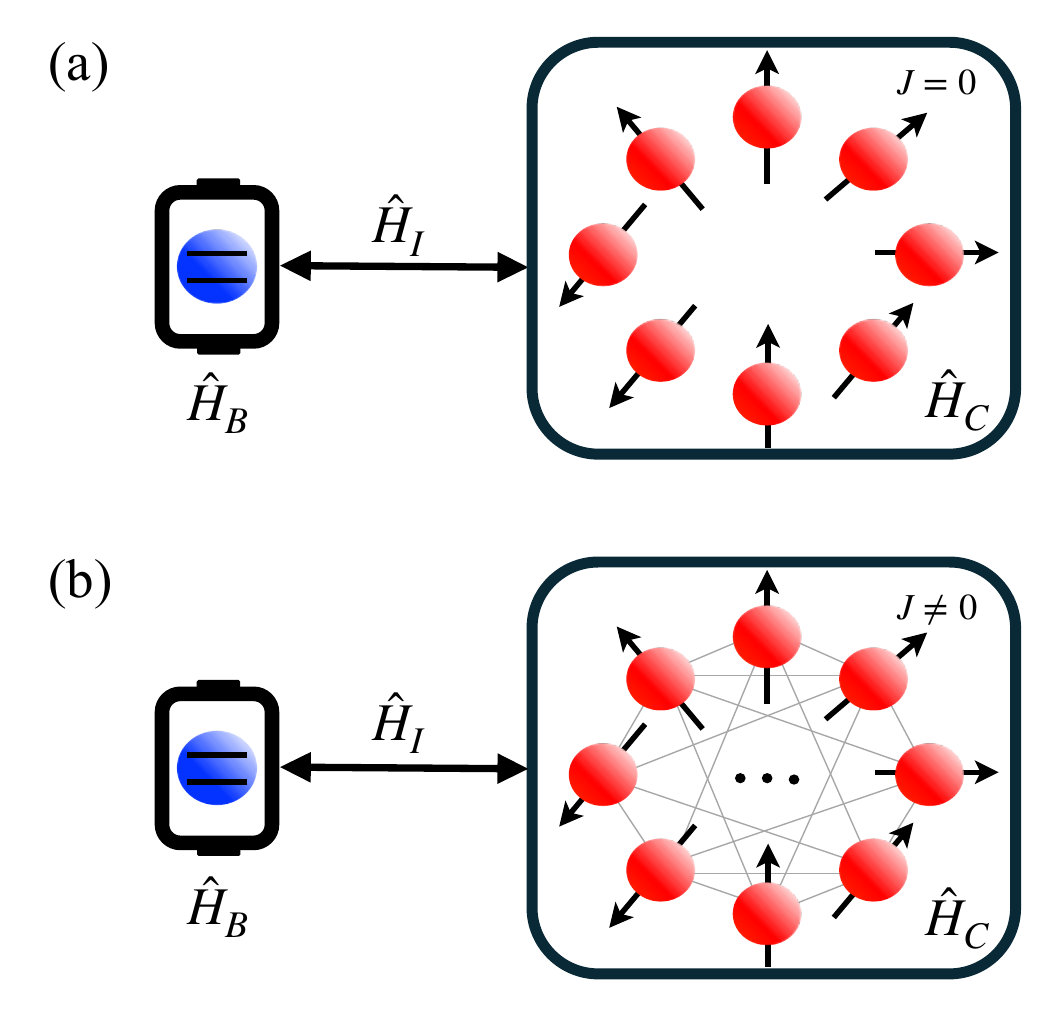}
    \caption{Schematic of a central qubit battery coupled to a many-body spin charger. (a) Non-interacting charger composed of \(N\) spin-\(1/2\) constituents. (b) Interacting charger with collective all-to-all spin interactions.}
    \label{fig:Fig1}
\end{figure}

We consider the charging of a single qubit quantum battery using a many-body quantum system. The battery comprises a simple two-level system,
\begin{equation}
    \hat{H}_B = \frac{\omega_B}{2}\hat\sigma^z\;,
    \label{eq:Hbattery}
\end{equation}
with \(\omega_B\) the battery energy-splitting and energy basis \(\left\{\ket{0}_B,\ket{1}_B\right\}\) and eigenvalues  \(\left\{-\omega_B/2,\omega_B/2\right\}\). The charger Hamiltonian is described by the LMG model, 
\begin{equation}
    \label{eq:Hcharger}
    \hat{H}_C = \omega_C \hat{S}_z - \frac{J}{N} \hat{S}_x^2\;,
\end{equation}
where \(\omega_C\) is the transverse field, \(J\) is the spin-coupling strength, and we use the collective-spin representation \(\hat{S}_\alpha \!=\! \frac{1}{2}\sum_{i\!=\!1}^N \hat{\sigma}_i^\alpha\) where \(\alpha \!\in\! \{x,y,z\}\). We restrict throughout to the fully symmetric collective-spin sector with total spin \(S\!=\!N/2\). When \(J\!=\!0\), the charger consists of \(N\) independent spins, whereas \(J\!\neq\!0\) couples spins collectively through an all-to-all interaction, defining the two battery--charger configurations considered here, schematically illustrated in Fig.~\ref{fig:Fig1}. In the thermodynamic limit \(N\!\to\!\infty\), the non-interacting LMG model can be mapped onto an effective quantum harmonic oscillator~\cite{Dusuel2004}, whose charging capabilities have been described in Ref.~\cite{Andolina2018}. By contrast, at finite \(N\), the charger retains a bounded and interaction dependent many-body spectrum, allowing us to explore how detuning, collective effects, and intra-charger interactions influence the spectrum and shape the charging process. 

The transfer of energy between the battery and charger is facilitated by the interaction,
\begin{equation}
    \label{eq:Hinteraction}
    \hat{H}_I = g\left(\hat{\sigma}_+ \otimes \hat{S}_- + \hat{\sigma}_- \otimes \hat{S}_+\right) \;,
\end{equation}
where \(\hat{S}_{\pm}=\hat{S}_x\pm i\hat{S}_y\) are the collective spin raising and lowering operators of the charger, respectively. \(\hat{H}_I\) exchanges a single excitation between the battery and charger, with \(g\) denoting the battery--charger coupling. As is standard, we consider the limit of weak battery--charger interaction, \(g\!\ll\! \omega_{B},\,\omega_C\), while allowing a range of intra-charger interaction strengths, \(J\). To charge the battery, we suddenly switch on the interaction at \(t\!=\!0\) and switch it off at time \(t\) to isolate the battery.

An idealized charging process would transfer energy to the battery without requiring an external energetic contribution associated with the coupling or decoupling the battery and charger. This can be achieved by enforcing the condition \( [\hat{H}_B \!+\! \hat{H}_C, \hat{H}_I ] \!=\! 0\) where an increase of the battery energy is supplied entirely by the charger. Whether this energy is stored as extractable work depends additionally on the resulting battery state. However, when \( [ \hat{H}_B + \hat{H}_C, \hat{H}_I ] \!\neq\! 0\), ambiguities in the origin of the extractable work arise.

Throughout, we consider initial states such that $\langle \hat H_I \rangle_{t=0}\!=\!0$, and therefore the switch-on work vanishes. 
During the charging process, however, energy can be redistributed into the interaction channel such that $\langle \hat H_I \rangle_{t>0}\neq 0$. Since the battery must ultimately be isolated, a nonzero interaction energy implies an energetic cost or yield upon uncoupling the battery and charger, corresponding to the work associated with switching off \(\hat H_I\). Consequently, the charging power and/or battery energy storage capabilities may be enhanced or deteriorated due to the energetics associated with the interactions. The extent to which the present charging protocol departs from the idealized condition is therefore encoded in the commutator:
\begin{equation}
\begin{aligned}
[\hat{H_B} + \hat{H}_C, \hat{H}_I]
&=
g\Delta\left(\hat{\sigma}_+ \otimes \hat{S}_- - \hat{\sigma}_- \otimes \hat{S}_+\right)
\\
&\quad
-\frac{gJ}{N}\,\hat{\sigma}_+ \otimes (\hat{S}_x \hat{S}_z + \hat{S}_z \hat{S}_x)
\\
&\quad
+\frac{gJ}{N}\,\hat{\sigma}_- \otimes (\hat{S}_x \hat{S}_z + \hat{S}_z \hat{S}_x),
\end{aligned}
\label{eq:EnergyCommutator}
\end{equation}
with \(\Delta \!=\! \omega_B - \omega_C\) the detuning. In the non-interacting charger regime, \(J\!=\!0\), the commutator vanishes only on resonance, providing a natural reference regime for ideal charging. Away from resonance, \(\Delta\!\neq\!0\), the spectral mismatch between the battery gap and charger transition frequency introduces an energetic contribution from the battery-charger interaction such that the energy deposited in the battery can no longer be attributed solely to the charger. Detuning in the non-interacting regime therefore provides a clean benchmark for isolating how spectral alignment controls the charging dynamics. In the interacting regime, \(J\!\neq\!0\), the commutator is non-zero, even when the battery is tuned to resonance with a dressed charger transition. Consequently, intra-charger interactions generically introduce an energetic contribution. We refer to this interaction-induced diversion of otherwise available charger energy away from direct battery charging as an energetic impedance. The term is used by analogy with electrical impedance: the suppression of direct transfer need not arise from dissipation, but can instead result from energy being coherently stored in and released through additional energetic channels. Any enhancement or deterioration of the charging performance must therefore be interpreted by separating the effects of this spectral restructuring from the energetic contribution of the battery-charger interaction itself.


\section{The non-interacting regime: \(J\!=\!0\)}
\label{sec:noninteracting_regime}

\subsection{Fock state}
\label{sec:noninteracting_regimeA}

\begin{figure*}[t]
    \centering
    \includegraphics[width=1\linewidth]{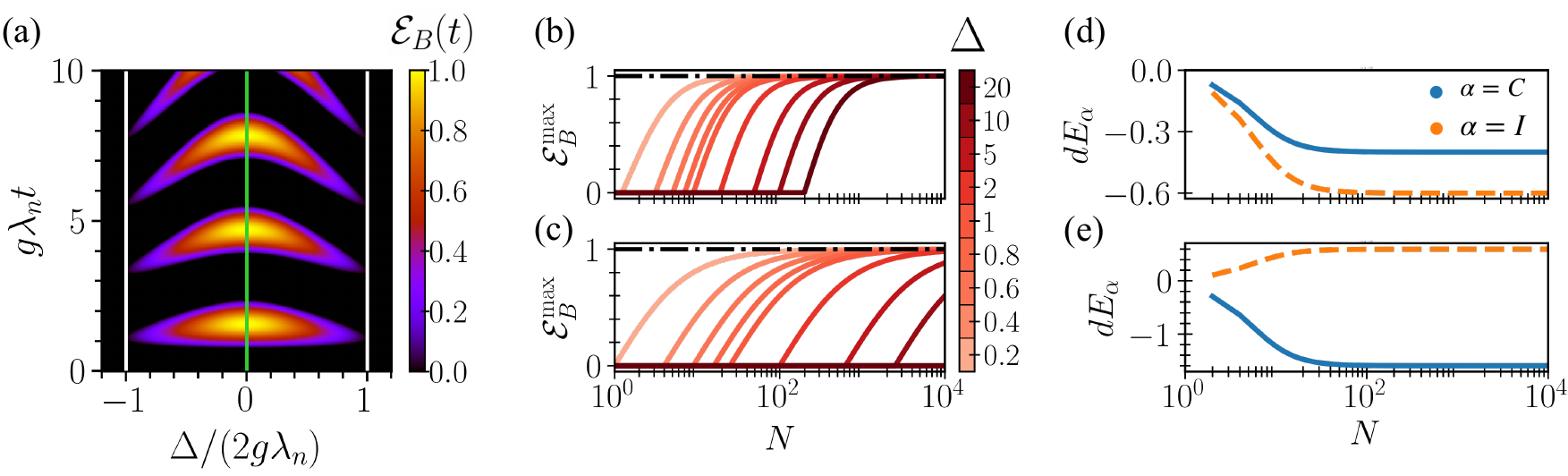}
    \caption{Panel (a): Ergotropy of the battery, \(\mathcal{E}_B(t)\), as a function of rescaled time (\(g\lambda_nt\)) and rescaled detuning (\(\Delta/2g\lambda_n\)). The white lines are the boundary defined in Eq.~\eqref{eq:window}. On resonance (green line) the battery gets fully charged, while off resonance, the battery ergotropy progressively decreases. Panels (b) and (c): Maximum battery ergotropy, \(\mathcal{E}_B^{max}\), as a function of charger size \(N\) for \(n\sim N/2\) in panel (b) and \(n\!=\!1\) for panel (c). We use  detuning values \(|\Delta| \!=\! 0.2,0.4,0.6,0.8,2,5,10,20\) with colour scheme light to dark for weak to strong detuning. Off resonance a minimum charger size is needed to activate the battery, with mid-spectrum states reaching this threshold at smaller \(N\) than spectral edge states. For fixed detuning in the thermodynamic limit the battery can become fully charged, $\mathcal{E}_B^{max}\to\omega_B$. Panel (d) and (e): Changes in charger (\(\alpha\!=\!C\)) and interaction (\(\alpha\!=\!I\)) energies for mid-spectrum state \(n\sim N/2\) as function of charger size \(N\), for detuning value \(\Delta\!=\!0.6\) in (d) and \(\Delta\!=\!-0.6\) in (e). The sign of detuning determines how the energetic balance is partitioned between the charger and interaction contributions. In panels (b)-(e) the maximum battery ergotropy, \(\mathcal{E}_B^{max}\), and $d E_{\alpha}$ are determined at the first local maximum of battery ergotropy, \(t\!=\!t_*\). Other parameters are \(g\!=\!0.1\) and \(\omega_B\!=\!1\).}
    \label{fig:Fig2}
\end{figure*}

In the non-interacting regime, \(J\!=\!0\), the charger Hamiltonian and exchange interaction conserve the total excitation number. Thus, for a charger prepared in the \(n\)-th Fock state and the battery in its ground state, the dynamics is restricted to the two-dimensional subspace \(\vert n \rangle_C \vert 0 \rangle_B\!\leftrightarrow\! \vert n\!-\!1\rangle_C\vert 1 \rangle_B\). Projecting onto this subspace gives,
\begin{equation}
    \label{eq:non-interacting_ham}
    \hat{H}_n = n \omega_C\mathbb{I}_2 - \frac{\Delta}{2}\hat{\sigma}_z + g \lambda_n \hat{\sigma}_x\;,
\end{equation}
where $\mathbb{I}_2$ is the identity matrix, \(\Delta\!=\!\omega_B\!-\!\omega_C\) is the battery--charger detuning and, 
\begin{equation}
    \lambda_n = \langle n-1 | \hat{S}_- | n \rangle = \sqrt{n(N-n+1)}\;,
    \label{eq:lambda}
\end{equation}
is the effective coupling matrix element, where \(n\,\in\,\{0,1, \dots N\}\) and \(N\) is the number of charger spins.  Eq.~\eqref{eq:lambda} sets the characteristic rate of excitation transfer, while its magnitude relative to the detuning, \(|\Delta|/(2g\lambda_n)\), determines the sensitivity of the charging dynamics to detuning. The ergotropy stored in the battery follows by directly computing Eq.~\eqref{eq:Ergotropy},
\begin{equation}
\label{eq:battery_ergotropy}
    \mathcal E^{(n)}_B(t) = \omega_B \text{max}[0,2P_n(t)-1]\;,
\end{equation}
where \(n\) denotes the chosen Fock state of the charger and \(P_n(t) \!=\! (g^2\lambda_n^2/\Omega_n^2) \sin^2{(\Omega_n t)}\) is the charging probability with \(\Omega_n\!=\!\sqrt{g^2 \lambda^2_n \!+\!\Delta^2/4}\) a generalized Rabi frequency.  In Fig.~\ref{fig:Fig2}(a) we show the battery ergotropy as a function of rescaled detuning \(\Delta/(2g\lambda_n)\) and time \(g\lambda_n t\) where the dynamics collapse onto a single curve, removing the explicit dependence on system size \(N\) and choice of initial Fock state \(n\). Clearly, we find that ideal charging occurs on resonance when \(\Delta\!=\!0\), denoted by the green vertical line. When the battery and charger are off resonance, \(\Delta\!\neq\!0\), we can determine a condition for the battery to store a finite amount of ergotropy by enforcing \(P_n(t)\!>\!1/2\). This defines a finite-detuning window which corresponds to a (partially) charged battery,  
\begin{equation}
    -2 g \lambda_n < \Delta < 2 g \lambda_n\;.
    \label{eq:window}
\end{equation}
Hence, away from resonance the ergotropy remains finite but decreases non-monotonically within the bounded region \(\vert\Delta/(2g\lambda_n)\vert\!< 1\), denoted by the solid white lines. Importantly, however, within this window ergotropy in the battery cannot be attributed exclusively to having come from the charger: for \(\Delta\!\neq\!0\), Eq.~\eqref{eq:EnergyCommutator} is strictly non-zero, allowing the interaction energy to contribute directly to the battery charging.

To unpack the roles of the charger size and initial Fock state, in Figs.~\ref{fig:Fig2}(b,c) we plot the maximum battery ergotropy as a function of system size \(N\), comparing a mid-spectrum state \(n\!=\!N/2\) in (b), with a spectral-edge state \(n\!=\!1\) shown in (c). On resonance, \(\Delta\!=\!0\), complete population inversion is always possible and \(\mathcal E^{\text{max}}_B\!=\!\omega_B\), independently of \(N\) for \(n\!\geq\!1\), although the charging time remains controlled by \(g\lambda_n\). Interestingly, off-resonance \(\Delta\!\neq\!0\), we see the emergence of a minimum system size below which the battery remains passive and no ergotropy is stored. This is because the detuning imposes an effective longitudinal field which tilts the rotation axis of \(\hat{H}_I\), competing with the effective coupling. The battery becomes active, and hence possesses finite ergotropy, only when, 
\begin{equation}
    N > N_c = n-1 + \frac{\Delta^2}{4g^2n}\;,
    \label{eq:minimum_N}
\end{equation}
valid for \(n\!\geq\!1\). This threshold follows directly from Eq.~\eqref{eq:window} and therefore depends on the choice of initial Fock state through \(\lambda_n\), where mid-spectrum states possess the largest transition matrix element, \(\lambda_{N/2}\!\simeq\!N/2\), compared to the spectral edge \(\lambda_1\!=\!\sqrt{N}\). Therefore, mid-spectrum states are capable of overcoming detuning at substantially smaller \(N\) and approach maximal ergotropy more rapidly, while increasing detuning shifts the onset of useful work to progressively larger charger sizes. Similar optimization of charger excitation sectors has recently been identified in central-spin quantum batteries, where distinguished sectors maximize charging performance~\cite{ShiPRB2025,ShiPRB2026}.

In Figs.~\ref{fig:Fig2} (b,c) all curves approach \(\mathcal E_B^{\text{max}}\!\to\!\omega_B\) in the large-\(N\) limit. This is because, for any fixed detuning, the collective coupling eventually dominates, \(g\lambda_n\!\gg\!\vert \Delta \vert\), rendering the effect of detuning asymptotically negligible. Nevertheless, complete battery charging away from resonance does not have the same energetic interpretation as resonant charging. Defining the average energies of the battery, charger, and interaction as,
\begin{equation}
    E_\alpha(t)=\langle H_\alpha\rangle = \operatorname{Tr}[\rho(t)H_\alpha],\qquad \alpha\in\{B,C,I\}
\end{equation}
with associated energy changes, \(dE_\alpha\!=\!E_\alpha(t)-E_\alpha(0)\). While the total energy remains conserved, the individual contributions satisfy \(d E_B \!+\! d E_C\!=\!-d E_I\), meaning the energy gained by the battery in general does not originate solely from the charger. 

To resolve the energetic origin of the energy deposited in the battery away from resonance, we now track the energy exchanged between the battery, charger, and interaction channels. This energetic accounting identifies the relevant sources and sinks of energy, but should not be conflated with the battery ergotropy. Indeed, comparing Eq.~\eqref{eq:battery_ergotropy} with the change in battery energy we find that the battery may acquire energy while remaining passive; only once the population becomes inverted does this stored energy become locally extractable as work and Eq.~\eqref{eq:battery_ergotropy} becomes non-zero. Without loss of generality, we fix the energy reference such that the zero-excitation state has zero energy. The sector-resolved average energies then follow as,
\begin{align}
E_B^{(n)}(t)
&=
\omega_B P_n(t),
\label{eq:EBm}\\
E_C^{(n)}(t)
&=
\omega_C\left[n-P_n(t)\right],
\label{eq:ECm}\\
E_I^{(n)}(t)
&=
-\Delta P_n(t).
\label{eq:Eintm}
\end{align}
Relative to the initial state, energy conservation requires
\(d E_B^{(n)}+d E_C^{(n)}+d E_I^{(n)}\!=\!0\), with
\(d E_I^{(n)}\!=\!-\Delta P_n(t)\). The interaction therefore acts as an
energetic source for positive detuning and as a sink for negative detuning,
as illustrated in Figs.~\ref{fig:Fig2}(d,e). For \(\Delta\!>\!0\), the charger
and interaction jointly supply the energy stored in the battery, whereas for
\(\Delta\!<\!0\), the charger releases more energy than the battery stores and
the excess is absorbed by the interaction. Crucially, this interaction-energy
contribution remains finite even as increasing \(N\) restores maximal battery
ergotropy. Thus, the thermodynamic limit removes the dynamical suppression
caused by detuning, but not its energetic consequence: away from resonance,
the battery energy cannot be attributed solely to the charger.


\subsection{Thermal states and the role of temperature}

The isolated Fock-state analysis establishes two features that are important to the following analysis where we will consider a thermal charger. Firstly, the charging dynamics strongly depend on which parts of the charger spectrum are occupied through Eq.~\eqref{eq:lambda}. Secondly, away from resonance, the energetic bookkeeping becomes nontrivial, as changes in the interaction energy can contribute directly to the energy acquired by the battery, i.e. the charger ceases to be the sole source of the ergotropy stored in the battery. To isolate the role of temperature from this latter effect, we focus exclusively on resonance (\(\Delta\!=\!0\)) for which \(\langle \hat H_I \rangle\!=\!0\) throughout the evolution. This provides a controlled setting in which to explore whether two locally passive subsystems, prepared out of {\it global} equilibrium such that their joint state is non-passive, can generate {\it local} ergotropy in the battery. To investigate finite-temperature effects, we initialise the charger in a Gibbs state. To maintain consistency with the interacting case discussed in Sec.~\ref{sec:interacting_regime}, we restrict throughout to the collective \(S\!=\!N/2\) manifold and therefore the Gibbs state is given by
\begin{equation}
    \hat \varrho_C^{(\beta)} = Z^{-1} \sum_{n=0}^N e^{-\beta \omega_C n} \vert n \rangle \langle n \vert\;,
\end{equation}
where $\lvert n\rangle$ denotes the same collective Fock ladder considered above and \(Z\!=\!\sum_{n\!=\!0}^N e^{-\beta \omega_C n}\) is the partition function. Since the charger has a bounded spectrum, we also consider negative inverse temperatures, \(\beta\!<\!0\), corresponding to initially active charger states. 

The battery qubit is again assumed to be initialized in its ground state, \(\ket{0}_B\).
As the thermal state is diagonal in the Fock basis, the battery excited state population can be decomposed as,
\begin{equation}
    p_B(t) = \sum_{n=0}^N p_n \sin^2{(g \lambda_n t)}
\end{equation}
where \(p_n\!=\!Z^{-1}e^{-\beta \omega_C n}\) and finite ergotropy is generated whenever \(p_B(t)\!>\!1/2\). Clearly the battery will gain energy as a result of the coupling to the charger, however, whether \(p_B(t)\) exceeds the threshold necessary for ergotropy to be present in the battery is not immediately obvious. Hence energy transfer alone is insufficient, and the central question of this section is to assess whether energy available in a (passive) thermal state of the charger can be accessed to produce a population-inverted battery.

\begin{figure}[t]
    \centering
    \includegraphics[width=1\linewidth]{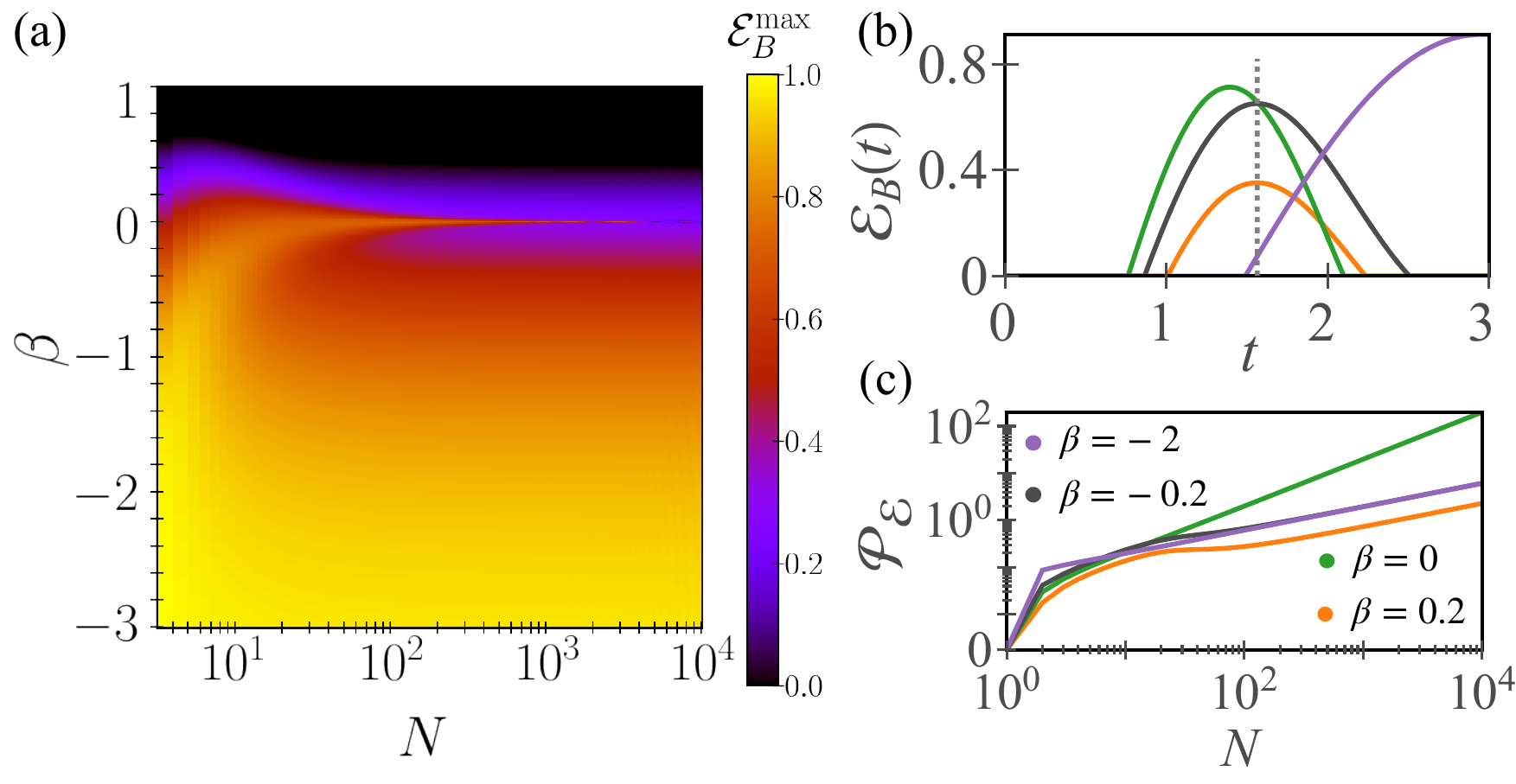}
    \caption{Panel (a): Maximum battery ergotropy, \(\mathcal{E}_B^{max}\), as function of charger inverse temperature \(\beta\) and size \(N\). Finite ergotropy reaches the battery once the thermal charger becomes sufficiently hot and large.  Panel (b): Battery ergotropy as a function of time, \(\mathcal{E}_B(t)\), for four different inverse temperatures of the charger \(\beta\!=\!0.2,\,0,\,-0.2,\,-2\). The vertical dashed line marks the common time of the first local maximum of $\mathcal{E}_B(t)$ for the \(\beta\!=\!\pm0.2\) pair. 
    \(N\!=\!25\).
    Panel (c): Ergotropic power, \(\mathcal{P}_{\mathcal{E}}(t_*)\), where \(t_*\) is the time of first local maximum of battery ergotropy for a given \(\beta\), as a function of charger size \(N\). We consider four different inverse temperatures of the charger \(\beta\!=\!0.2,\,0,\,-0.2,\,-2\), with the same colour scheme as panel (b). For the finite-positive-temperature and infinite-temperature regimes the ergotropic power scales as \(\sim N^{1/2}\) and \(N\) respectively. This arises from the scaling of \(\lambda_n\). \(g\!=\!0.1\), and \(\Delta\!=\!0\) for all panels, such that the mean interaction energy vanishes and the battery-energy gain is supplied by the charger.}
    \label{fig:Fig3}
\end{figure}

Fig.~\ref{fig:Fig3}(a) illustrates precisely that finite battery ergotropy can be generated using an initially passive charger state. Explicitly, when \(\beta\!\geq\!0\), the charger is completely passive and contains no initial ergotropy. Nevertheless, finite battery ergotropy can be achieved provided the charger is sufficiently large and is at a high temperature. This is a direct consequence of the spectral structure identified in Sec.~\ref{sec:noninteracting_regimeA}. In particular, at low temperature the Gibbs distribution becomes concentrated near states at the lower spectral edge where the populated sectors have a small transition rate \(\lambda_n\). An increase in temperature correspondingly results in a spread of the charger population into the mid-spectrum states, where \(\lambda_n\) becomes larger---allowing a greater fraction of thermal energy to effectively participate in the charging process. Consequently, the charger size \(N\) and inverse temperature \(\beta\) complement one another: an increase in \(N\) enlarges the number of available collective matrix elements, whereas an increase in temperature allows the corresponding mid-spectrum sectors to become more accessible. 

The underlying spectral picture explains the observed charging dynamics in Fig.~\ref{fig:Fig3}(b,c), where we show the ergotropy, \(\mathcal{E}_B(t)\) for several temperatures and the ergotropic power, defined as,
\begin{equation}
    \mathcal{P}_{\mathcal{E}}(t) = \frac{\mathcal{E}_B(t)}{t}.
\end{equation}
At any fixed finite positive temperature that is sufficient to charge the battery, the Gibbs weight remains concentrated within \(n\!=\!\mathcal O(1)\) levels of the lower spectral edge as \(N\) grows, for which \(\lambda_n\!\sim\!\sqrt N\). The characteristic charging time, \(t_*\), defined as the first local maximum of battery ergotropy,  scales as \(t_* \!\sim\!N^{-1/2}\) yielding an ergotropic power \(\mathcal P_E(t_*)\!\sim\!\sqrt N\). This is in direct contrast to the infinite temperature regime, where all sectors are initially equally populated and substantial Gibbs weight lies across the mid-spectrum states, where \(\lambda_n\!\sim\!N\). This results in faster scaling \(t_*\!\sim\!N^{-1}\) and \(\mathcal P_E(t_*)\!\sim\!N\). Therefore, even when the maximum ergotropy saturates, increasing the charger size can strongly enhance the rate at which thermal energy contributes to storing extractable work in the battery. Negative temperature states, $\beta\!<\!0$, correspond to population inverted, initially active charger states and yield the largest ergotropy, as shown in the lower portion of panel (a), albeit at a slower rate, as reflected in the ergotropy and ergotropic power for $\beta\!=\!-0.2$ and $-2$ shown in panels (b) and (c) respectively.
 
A simple necessary condition for battery activation follows from the fact that the charger ground state cannot transfer an excitation to the battery. Since only the excited charger sectors can contribute to the battery excitation, the maximum achievable excited-state population is bounded by \(p_B(t)\!\leq\! 1-p_0\). Population inversion, and hence finite battery ergotropy, therefore requires, \(p_0\!<\!1/2\). For the finite-\(N\) thermal distribution, \(p_0\!=\!\frac{1-q}{1-q^{N+1}}
\), where \(q\!=\!e^{-\beta\omega_C}\). This provides a finite-size activation threshold. In the thermodynamic limit \(N\rightarrow\infty\), for positive temperature, \(q\!<\!1\), and hence \(p_0\rightarrow 1-q\). The activation condition then reduces to \(q\!>\!1/2\), or equivalently \(\beta\omega_C\!<\!\ln 2.\)
Using \(\langle n\rangle\!=\!q/(1-q)\), this may be written in the simple form \(\langle n\rangle\!>\!1.\) It follows then that in the thermodynamic limit the charger must contain on average more than one excitation for battery activation to be possible. 

\begin{figure}[t]
    \centering
    \includegraphics[width=\linewidth]{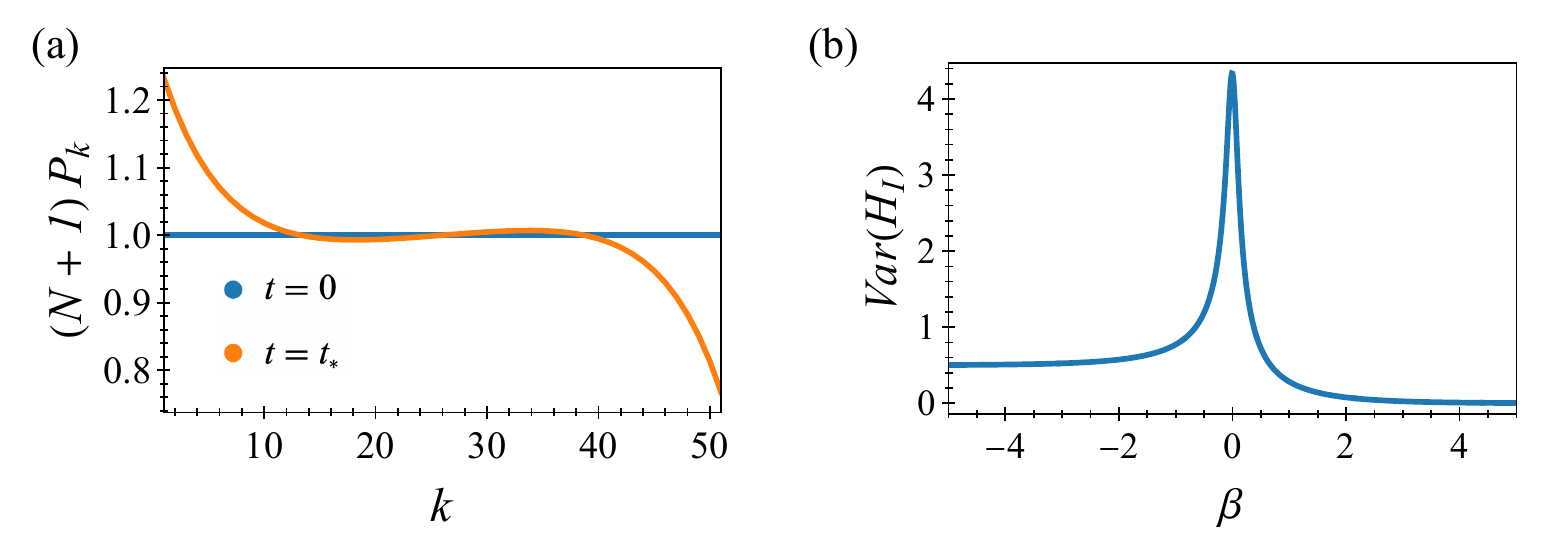}
    \caption{Panel (a): To show the restructuring of the charger's state during charging we calculate the rescaled population \((N+1)P_k\) of the \(k\)th charger eigenstate at the initial time \(t\!=\!0\) and at the first maximum of the battery ergotropy, \(t\!=\!t_*\), for an infinite-temperature charger, \(\beta\!=\!0\). Panel (a) highlights the redistribution of population from the upper part of the charger spectrum towards lower-energy states.
    Panel (b): Variance of the interaction, \(\text{Var}(\hat{H}_I)\), as a function of the inverse temperature \(\beta\). This highlights that while the average interaction energy is zero, the variance of the interaction Hamiltonian is non-trivial. \(N\!=\!50, \, g\!=\!0.1, \, \text{and } \Delta\!=\!0\) for both panels.}
    \label{fig:Fig4}
\end{figure}

The restructuring of the charger's population distribution responsible for this transfer is visible directly in the charger populations in Fig.~\ref{fig:Fig4}(a). During charging, population is shifted toward lower charger energies as the battery becomes excited. On resonance, \([\hat H,\hat H_I]=0\) holds, meaning both the mean and variance of the interaction energy are conserved throughout the charging dynamics. While the mean interaction energy vanishes, making the average energetic accounting simple, its non-zero fluctuations are given by,
\begin{equation}
    \text{Var}[\hat H_I]_\beta = g^2 \sum_{n=0}^N p_n \lambda_n^2\;,
\end{equation}
which directly weights the same state dependent matrix elements governing the charging dynamics. Fig.~\ref{fig:Fig4}(b) shows the interaction energy fluctuations are largest near infinite temperature, where thermal states sample the strongly coupled mid-spectrum states of the charger. This connection can be made explicit for \(\beta\!=\!0\) as \(\text{Var}[\hat H_I]_{\beta\!=\!0} \!\sim\! N^2\). For any fixed positive temperature, the Gibbs distribution remains dominated by edge states and the fluctuations are given by \(\text{Var}[\hat H_I]_\beta \!\sim\!N\). The same crossover from edge scaling to bulk scaling underlies the enhancements of the charging power in Fig.~\ref{fig:Fig3}. 

Thus, the resulting picture is more nuanced than the vanishing mean interaction energy alone would suggest. On resonance, the interaction supplies no net energy to the battery, however, the interaction possesses substantial energy fluctuations whose magnitude is controlled by the parts of the spectrum the many-body charger thermally occupies. These fluctuations do not constitute an additional energy source on average. Instead, the fluctuations quantify the dynamical coupling available to mediate the conversion of passive thermal energy into locally extractable work---whose effectiveness is determined jointly by its thermal population and the spectral structure of its exchange matrix elements.



\section{The interacting regime: \(J\!\neq\!0\)}
\label{sec:interacting_regime}

\subsection{Eigenstate}
\label{sec:eigenstate_interacting}
We now turn to the interacting charger, \(J\!\neq\!0\). In contrast to the non-interacting case, the charger eigenstates are no longer single Fock states. Instead, the quadratic interaction, \(\hat S_x^2\), dresses sectors differing by two excitations. Consequently, the battery--charger exchange can access several many-body transitions previously unavailable. The charging dynamics are therefore no longer restricted to a single two-dimensional subspace, now requiring a numerical treatment. However, this makes the interacting problem considerably richer. Indeed, the strength of the interaction, \(J\), not only changes the structure of the internally stored energy and spectrum, but sufficiently strong interactions can probe the non-perturbative and critical structure of the underlying LMG charger. 
We restrict our discussion to \(\Delta\!=\!0\) such that the battery is resonant with the bare single-particle charger frequency. For $J\!\neq\!0$, the relevant dressed transition frequency is generally shifted. We quantify the shift below with the goal of isolating how the intra-charger interaction modifies the ability to store ergotropy in the battery.

We first consider the weak-\(J\) limit. For the \(n\)-th dressed charger eigenstates, first order perturbation theory gives the transition frequency,
\(\omega_{(n,n-1)}(J) \simeq \omega_C + \frac{J}{2N}(2n-N-1)\),
with corresponding battery--charger detuning,
\begin{equation}
    \Delta_n^{\text{eff}}(J) \simeq \Delta + \frac{J}{2N}(N+1-2n)\;,
\end{equation}
shifting the relevant resonance away from the bare condition, \(\Delta\!=\!0\). Here, the corresponding exchange matrix element remains \(M_{n-1,n}(J) \!=\! \lambda_n + \mathcal O(J^2)\)---the internal interaction does not directly modify the strength of the dominant charging channel, but shifts the transition frequency. In the weakly interacting regime, the \(J\)-dependent shift of the charger transition frequency plays a role analogous to detuning in the non-interacting case of Sec.~\ref{sec:noninteracting_regimeA}, leading to the finite-ergotropy window,
\begin{equation}
    \vert \Delta + \frac{J}{2N}(N+1-2n)\vert < 2g \lambda_n\;.
    \label{eq:perturbative_window}
\end{equation}

\begin{figure}[t]
    \centering
    \includegraphics[width=\linewidth]{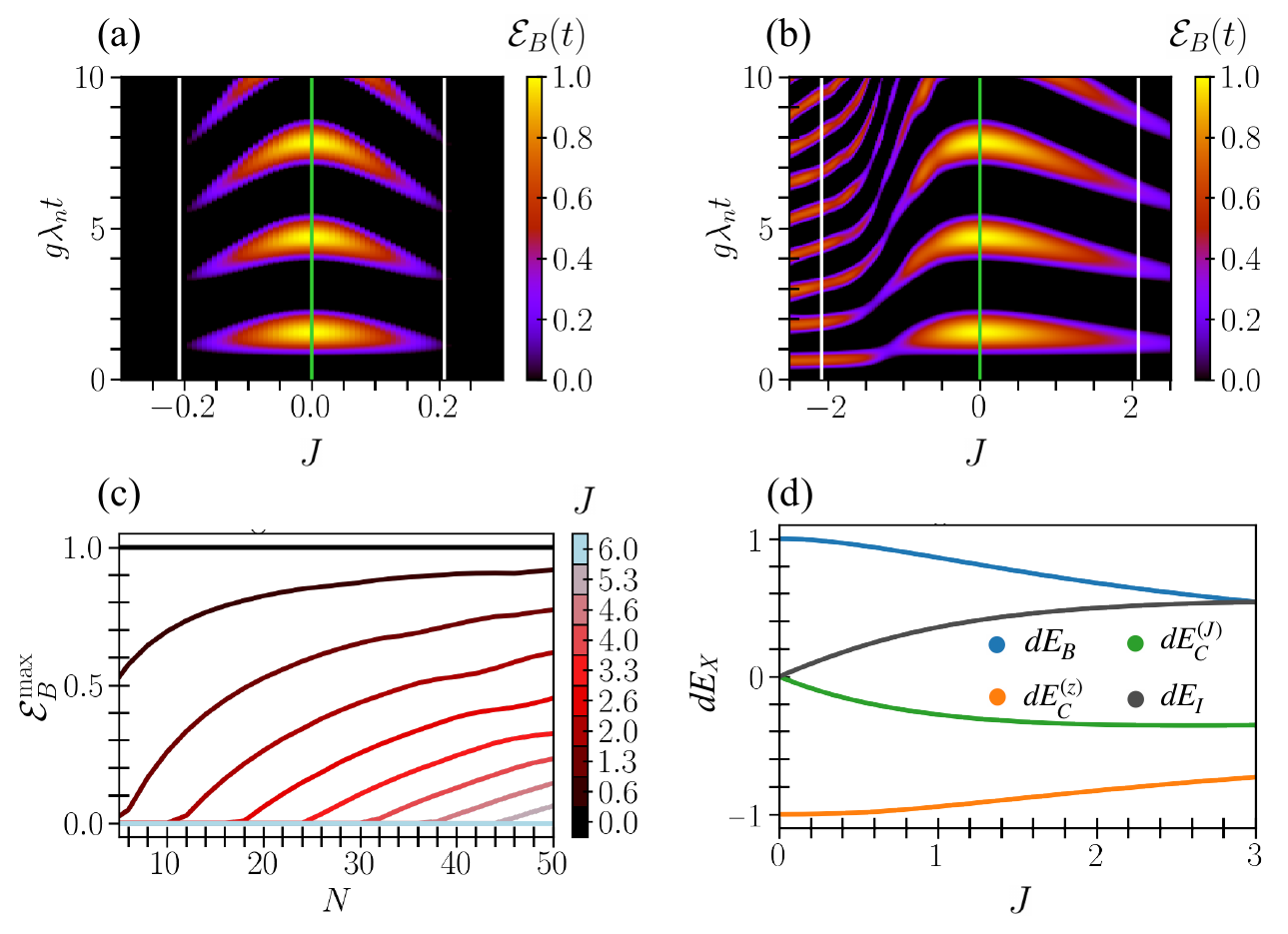}
    \caption{Panels (a) and (b): Battery ergotropy, $\mathcal E_B(t)$, as a function of charger interaction strength $J$ and rescaled time $g\lambda_n t$, for $g\!=\!0.01$ and $g\!=\!0.1$, respectively. The charger is initialised in the first excited state and \(N\!=\!25\). The white lines indicate the perturbative finite-ergotropy window of Eq.~\eqref{eq:perturbative_window}. As the battery-charger coupling strengthens, the perturbed two-level description breaks down. Panel (c): Maximum battery ergotropy, \(\mathcal{E}_B^{max}\), when the charger is initialised in the first excited state, as a function of charger size $N$ for several values of $J$ and \(g\!=\!0.1\). As \(J\) increases a minimum charger size emerges in order to activate the battery. Panel (d): Changes in the bare battery energy, bare charger energy, interaction-induced charger energy, and battery--charger interaction energy, $dE_B$, $d E_C^{(z)}$, $d E_C^{(J)}$, and $d E_I$, respectively, evaluated at the time of maximum battery ergotropy as functions of $J$ for $N\!=\!25$ and \(g\!=\!0.1\). Increasing the intra-charger interaction strength produces an energetic impedance to direct charger-to-battery transfer, as an increasing fraction of the available charger energy is diverted into the interaction channel. \(\Delta\!=\!0\) for all panels.}
    \label{fig:Fig5}
\end{figure}

In Fig.~\ref{fig:Fig5}(a,b) we test the weak-\(J\) picture directly, plotting the battery ergotropy, \(\mathcal E_B(t)\) as a function of \(J\) and rescaled time \(g\lambda_n t\), where dashed lines mark the perturbative finite-ergotropy window Eq.~\eqref{eq:perturbative_window}. The two panels probe different battery--charger couplings, \(g\!=\!0.01\) and \(g\!=\!0.1\) respectively. For the weaker coupling, the allowed region is narrow and remains confined to \(J\!\ll\!1\), with the perturbative prediction following the exact dynamics. Increasing to \(g\!=\!0.1\) broadens the window into a regime in which the weak-\(J\) prediction is no longer valid. Here, the exact dynamics show additional resonance structures emerging as the charger spectrum becomes non-perturbatively reorganized. This shows that while \(J\) initially acts as a state-dependent detuning, stronger internal interactions require the full many-body description, where several dressed transition frequencies participate in the charging dynamics. 

Having established how the interaction modifies the relevant charging resonances, we now ask whether increasing the charger size can overcome the resulting spectral mismatch. In Fig.~\ref{fig:Fig5}(c) we plot the maximum ergotropy of the battery for a charger initialised in a spectral edge state, here chosen to be the first excited state, as a function of \(N\) for several values of \(J\). As established in Sec.~\ref{sec:noninteracting_regimeA}, complete population inversion is achievable for any charger size \(N\) in the resonant, non-interacting limit (\(J\!=\!\Delta\!=\!0\)). However, for finite \(J\), the edge state can remain below the inversion threshold over an extended range of charger sizes, yielding vanishing ergotropy before eventually recovering. This is a consequence of the collective exchange matrix element becoming sufficiently large. Increasing \(J\) progressively shifts this onset to larger \(N\). Within the weak \(J\) approximation, the corresponding asymptotic threshold scales as \(N_C\!\sim\!(J/4g)^2\), providing the interacting analogue of Eq.~\eqref{eq:minimum_N}. Physically, increasing \(N\) does not remove the interaction-induced shift of the charger transition; rather, the collective exchange strength, \(g\lambda_1\!=\!g \sqrt{N}\), eventually becomes large enough to overcome it. As in Fig.~\ref{fig:Fig2}(b,c), this demonstrates that increasing the charger size can serve as a resource for overcoming the spectral mismatch, which in this case arises from the intra-charger interactions. 

The energetic accounting reveals an important distinction that is absent in the non-interacting charger. We separate the charger Hamiltonian as \(\hat H_C \!=\! \hat H_C^{(z)} \!+\! \hat H_C^{(J)}\) with \(\hat H_C^{(z)}\!=\! \omega_C \hat S_z\) and \(\hat H_C^{(J)}\!=\! \!-\! (J/N)\hat S_x^2\). Energy conservation then requires \(d E_B \!+\! d E_C^{(z)} \!+\! d E_C^{(J)} \!+\! d E_I \!=\! 0\). This decomposition allows us to resolve microscopically how much of the battery energy originates from the bare charger and from the internal many-body interaction. In Fig.~\ref{fig:Fig5}(d), we examine this balance at fixed system size, \(N\!=\!25\), as a function of coupling \(J\). At \(J\!=\!0\), the resonant result is recovered and the battery gain is supplied entirely by the bare \(\hat S_z\) sector, while \(d E_C^{(J)}\) and \(d E_I\) vanish. However, increasing \(J\) results in the internal structure \(\hat S_x^2\) becoming an important energetic contributor, with \(d E_C^{(z)}\) increased as the contribution \(d E_C^{(J)}\) decreases. We might therefore expect the additional internal interaction to act as a resource which improves the transfer of energy to the battery. Instead, we note the interaction energy \(d E_I\) becomes increasingly positive with growing \(J\), such that a growing portion of energy released by the charger becomes stored in \(\hat H_I\) rather than being deposited in the battery. Therefore, the internal interaction provides an additional energetic resource, but simultaneously increases the energetic participation of the battery--charger coupling required to access it. In this sense, \(J\) introduces an energetic impedance to direct charger-to-battery energy transfer: more internal charger energy becomes available, however, an increasing fraction of this energy is diverted into the interaction channel rather than to the battery. This ultimately diminishes the portion of the energy that can be stored as ergotropy in the battery. This behaviour is directly related to the non-vanishing commutator identified above: for \(J\neq0\), switching on \(\hat H_I\) introduces an additional interaction-energy channel, so the observed impedance reflects the fact that the coupling can no longer mediate a purely direct charger-to-battery energy exchange.

\subsection{Temperature}
The eigenstate analysis of Sec.~\ref{sec:eigenstate_interacting} provides insight into how intra-charger interactions reshape individual charging channels. We now consider a Gibbs-state charger, for which contributions from across the dressed spectrum become relevant. From the non-interacting case we have seen that temperature determines how the different excitation sectors are populated: equally spaced spectra are populated such that the charging process results from an incoherent average over the corresponding excitation resolved channels. At finite \(J\), however, this picture changes considerably. The Gibbs state now samples the interaction-dressed eigenstates according to their \(J\)-dependent energies, so that the intra-charger interaction modifies not only the subsequent charging dynamics, but also which many-body states are thermally available in the initial charger. In addition, the charger Hamiltonian itself naturally separates into single- and many-body interaction contributions \(\hat H_C \!=\! \hat H_C^{(z)} + \hat H_C^{(J)}\). In this sense, temperature and interaction strength are no longer energetically independent as the strength of \(J\) restructures the spectrum that the Gibbs distribution samples, while \(\beta\) determines which parts of the dressed spectrum participate in the charging process. 

\begin{figure}[t]
    \centering
    \includegraphics[width=\linewidth]{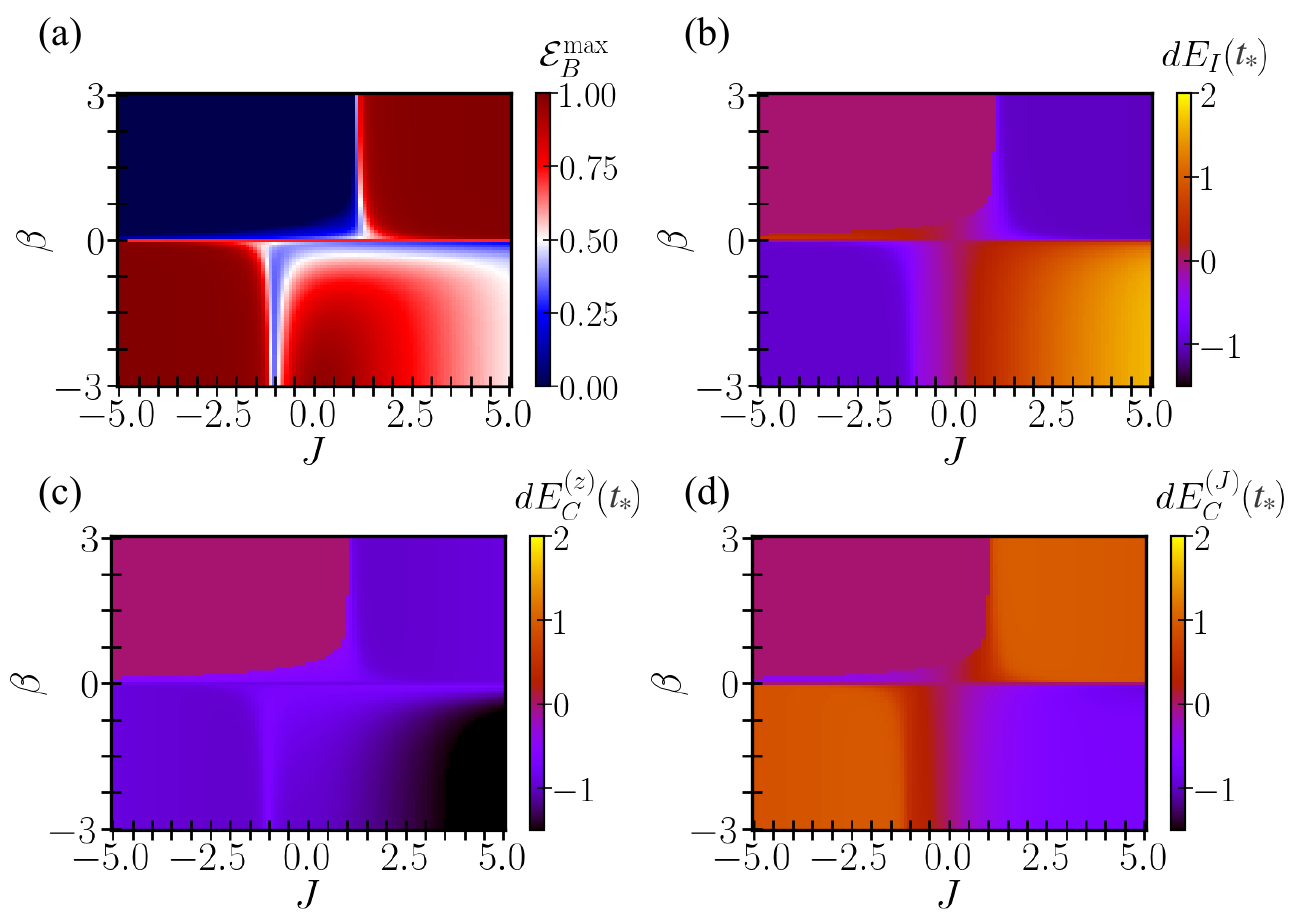}
    \caption{(a) Maximum battery ergotropy, $\mathcal E_B^{\max}$, as a function of inverse temperature $\beta$ and intra-charger interaction strength $J$. (b) Change in the battery--charger interaction energy, $d E_I$, evaluated at the time of maximum battery ergotropy. (c) Change in the single-particle charger contribution, $d E_C^{(z)}$, at the same time. (d) Change in the many-body charger contribution, $d E_C^{(J)}$, at the same time. Here $N=200$, \(\Delta=0\), and $g=0.1$.}
    \label{fig:Fig6}
\end{figure}

In Fig.~\ref{fig:Fig6}(a) we show the maximum battery ergotropy as a function of intra-charger interaction strength \(J\) and inverse temperature \(\beta\) for a fixed system size \(N\!=\!200\) for \(g\!=\!0.1\). Two pronounced high-ergotropy regions occur for \(\beta\!>\!0\), \(J\!>\!\omega_C\) and \(\beta\!<\!0\), \(J\!<\!-\omega_C\). The former case is particularly notable as the charger is initially a passive positive temperature Gibbs state. The origin of this enhanced charging, however, cannot be inferred from the ergotropy alone. Panels (b-d) resolve the corresponding energy changes, \(d E _I\), \(d E ^{(z)}_C\), and \(d E ^{(J)}_C\). In the two regions of highest ergotropy, the same qualitative energetic structure emerges. The single particle contribution is depleted \(d E_C^{(z)}\!<\!0\) whereas the many-body contribution increases \(d E_C^{(J)}\!>\!0\). Therefore, these two components of the charger compete with each other: energy released from the \(\hat S_z\) sector is partly stored in the intra-charger interaction rather than being transferred directly to the battery. This situation corresponds to \(d E_I\!<\!0\), so the battery--charger interaction releases energy and compensates for this internal energetic cost. The large ergotropy in (a) consequently results from a non-trivial energy balance whereby the bare charger contribution and the coupling act as energetic sources, while the many-body component acts as a sink. This reflects the fact that finite \(J\) restructures the charger spectrum and introduces an additional many-body energetic channel through \(\hat H_C^{(J)}\). However, the augmented charger structure associated with finite $J$, while giving rise to non-trivial many-body effects, does not necessarily translate into an energetic resource available for charging the battery. This is a reflection of the non-vanishing commutator in Eq.~\eqref{eq:EnergyCommutator} for finite $J$, which makes the interaction an energetically participating channel and prevents the large battery ergotropy from being interpreted as energy supplied directly by the many-body part of the charger.

Interestingly, the opposite competition also manifests itself. For example, when \(\beta\!<\!0\) and \(J\!>\!0\), the negative-temperature Gibbs state predominantly samples the upper part of the charger spectrum, which remains approximately equally spaced and largely dominated by the \(\hat S_z\) contribution even at strong \(J\). The battery therefore couples primarily to a less strongly correlated part of the spectrum, rather than to the quasi-degenerate, interaction-dominated low-energy sector that we associate with the energetic impedance. Accordingly, both charger contributions decrease, \(d E_C^{(z)}\!<\!0\) and \(d E_C^{(J)}\!<\!0\), and act as energetic sources, while \(d E_I\!>\!0\) indicates that the battery–charger interaction absorbs part of the released energy. Despite the greater depletion of the charger, the resulting ergotropy is not maximal. This contrast highlights that the energy released by the charger and the extractable work generated in the battery are distinct quantities. The total energetic balance is set by the combined changes in the charger and battery--charger interaction energies, while the battery ergotropy depends on how that available energy is redistributed into a locally active state. Consequently, a large depletion of the charger does not necessarily translate into large ergotropy being stored in the battery, since part of the released energy may be stored in the interaction channel or transferred in a form that does not produce sufficient population inversion in the battery.

Fig.~\ref{fig:Fig6} makes explicit the energetic redistribution in the charging process for the interacting LMG model. At \(J\!=\!0\) and \(\Delta\!=\!0\), the \(J\)-dependent charger contribution vanishes and the interaction channel does not participate energetically, recovering the transparent exchange discussed in the non-interacting case. For finite \(J\), however, we observe a strong dependence on both \(J\) and \(\beta\), with a substantial reduction in battery ergotropy in some regions, alongside a redistribution of energy between the \(\hat H_C^{(z)}\), \(\hat H_C^{(J)}\), and \(\hat H_I\) channels. The regions of maximal ergotropy therefore provide the clearest example of the central distinction of the interacting problem: many-body interactions can make the dynamics highly effective at charging the battery, while simultaneously introducing an energetic impedance to direct charger-to-battery energy transfer.
Charging performance and energetic provenance must therefore be assessed together.

\section{Conclusion}

In this work we asked what determines whether a finite many-body system can function as an effective quantum charger. Using the LMG model as a charger for a central-qubit battery allowed us to separate several ingredients involved in many-body charging: spectral alignment, collective enhancement, finite size, thermal occupation, and internal interactions. In the non-interacting regime, charging is governed by excitation-resolved channels with coupling strength \(\lambda_n\), while detuning limits population transfer and can impose a minimum charger size that depends on the initially occupied spectral sector. Increasing \(N\) overcomes this dynamical mismatch, but not the associated energetic contribution arising from the non-energy-conserving battery--charger exchange. For thermal states, the same spectral structure determines activation: despite both subsystems being locally passive, the battery can acquire ergotropy once sufficient population occupies excited charger sectors; in the thermodynamic limit, a necessary condition is simply \(\langle n\rangle\!>\!1\). More generally, this highlights that transferring energy to the battery and converting that energy into locally extractable work are distinct aspects of the charging process.

Internal interactions qualitatively modify both the charging dynamics and their energetic interpretation. For weak $J$, they primarily shift the charger transition frequencies, producing a state-dependent effective detuning, while at stronger coupling multiple dressed transitions participate and the two-level picture breaks down. At the same time, $J\neq0$ makes the battery--charger interaction generically non-energy-conserving with respect to the uncoupled Hamiltonians, such that energy released within the charger can be redistributed between its many-body interaction energy, the battery, and the battery--charger coupling. This gives rise to an energetic impedance to direct charger-to-battery transfer. Interactions can therefore generate spectral features that appear favourable for charging, but this apparent advantage is accompanied by a non-trivial energetic contribution associated with accessing them. For $J\!\neq\!0$, the non-vanishing commutator in Eq.~\eqref{eq:EnergyCommutator} makes the battery--charger coupling an energetically participating channel, such that interaction-induced spectral restructuring is accompanied by a redistribution of energy, including into the interaction channel, rather than translating directly into energy available for charging the battery. Consequently, the battery ergotropy alone does not reveal how much of the extractable energy originated solely from the charger. Assessing the effectiveness of a many-body charger therefore requires considering not only the useful work generated in the battery, but also the underlying energy flows and their energetic origin.

\acknowledgements
D.M., I.D’A., and S.C.  are grateful to the Royal Society International Exchanges Scheme IES$\backslash$R2$\backslash$242072. G.M. and S.C. acknowledge support from Taighde \'Eireann -- Research Ireland under grant number 24/EPSRC/4121. A.M. acknowledges funding by the Deutsche Forschungsgemeinschaft (DFG, German Research Foundation) – Project number 549314503. This publication has emanated from research conducted with the  support of Taighde \'Eireann -- Research Ireland, under Grant No.~23/RC/12197 at Rinn Quantum.



\bibliography{bibfile}

\end{document}